\documentclass[aps,prx,amsmath,amssymb,floatfix,twocolumn]{revtex4-1}
\usepackage{parskip} 
\usepackage{tikz} 
\usepackage{graphicx}
\usepackage{subfigure}
\usepackage{amsmath}
\usepackage{bbold}
\usepackage{slashed}
\usepackage{amssymb}
\usepackage{braket}
\usepackage{array}
\usepackage[colorlinks]{hyperref}
\usepackage{float}
\usepackage[margin=1in]{geometry}

\usepackage{natbib}

\begin{document}
\preprint{APS/123-QED}

\title{Electron-phonon coupling in copper-substituted lead phosphate apatite}
\author{Alexander C. Tyner$^{1,2}$}
\author{Sin\'ead M. Griffin$^{3,4}$}
\author{Alexander V. Balatsky$^{1,2}$}

\affiliation{$^{1}$ Nordita, KTH Royal Institute of Technology and Stockholm University 106 91 Stockholm, Sweden}
\affiliation{$^{2}$ Department of Physics, University of Connecticut, Storrs, Connecticut 06269, USA}
\affiliation{$^{3}$Materials Sciences Division, Lawrence Berkeley National Laboratory, Berkeley, California, 94720, USA}
\affiliation{$^{4}$Molecular Foundry Division, Lawrence Berkeley National Laboratory, Berkeley, California, 94720, USA}

\date{\today}

\begin{abstract} 
Recent reports of room-temperature, ambient pressure superconductivity in copper-substituted lead phosphate apatite, commonly referred to as LK99, have prompted numerous theoretical and experimental studies into its properties. As the electron-phonon interaction is a common mechanism for superconductivity, the electron-phonon coupling strength is an important quantity to compute for LK99. In this work, we compare the electron-phonon coupling strength among the proposed compositions of LK99. The results of our study are in alignment with the conclusion that LK99 is a candidate for low-temperature, not room-temperature, superconductivity if electron-phonon interaction is to serve as the mechanism.
\end{abstract}

\maketitle
\par 
\section{Introduction}
A solid-state system admitting room temperature, ambient pressure superconductivity is commonly regarded as a ``holy grail" of condensed matter physics due to the tremendous impact such a discovery would have on modern technology. In recent months, copper substituted lead apatite, commonly referred to as LK99, was proposed to support such a phenomena\cite{lee2023firs,lee2023superconductor}, prompting a widespread theoretical and experimental effort to better understand its properties.
\par 
A subsequent theoretical study computed the bulk band structure of LK99 considering two distinct locations for copper substitution, denoted Pb(1) and Pb(2), in the lead-phosphate apatite\cite{griffin2023origin}, Pb$_{10}$(PO$_{4}$)$_{6}$X$_{2}$, where X=O, OH. In each case, the existence of nearly-flat bands at the Fermi energy was demonstrated. The presence of such bands increases the density of states at the Fermi energy increasing the possibility of a mechanism for high-temperature superconductivity. 
\par 
In the intervening months many questions about the nature of electronic states in LK99 were raised. Additional studies have been published with remarkable speed and detail\cite{cabezas2023theoretical,hao2024first,jiang2308pb9cu,lai2024first,korotin2023electronic,lee2023effective,shen2023phase}, providing evidence that suggest LK99 is \emph{not} a high-temperature superconductor. {\color{black}It is important to note that despite being one of the most common mechanisms for superconductivity, the electron-phonon coupling in LK99 has \emph{not} been examined in detail for each molecular configuration. We are aware of only one work that has directly examined electron-phonon coupling strength\cite{paudyal2023implications}. This is expected given the computational demands of such a quantity as well as the fact that many available density functional theory packages do not offer the capability for computation of electron-phonon coupling strength if a non-zero value of the Hubbard U is specified for a constituent atom in the compound\cite{RevModPhys.89.015003}. Many works have instead chosen to examine alternative indicators of superconductivity which are computationally less expensive such as the Fubini-Studi metric\cite{jiang2308pb9cu}}. 
\par 
In Ref. \cite{paudyal2023implications}, the electron-phonon coupling strength was computed from first-principles for Pb$_{9}$Cu(PO$_{4}$)$_{6}$O$_{2}$ with Cu substituted at the Pb(2) site neglecting inclusion of a Hubbard U. It was shown that the electron-phonon coupling is weak. Even without inclusion of a Hubbard U for copper, this was a fascinating result as the electron-phonon interaction is one of the most common routes to the realization of superconductivity. It is thus a quantity of supreme importance and should be compared among the possible compositions of copper substituted lead phosphate apatite. 
\begin{figure}
    \centering
    \includegraphics[width=8cm]{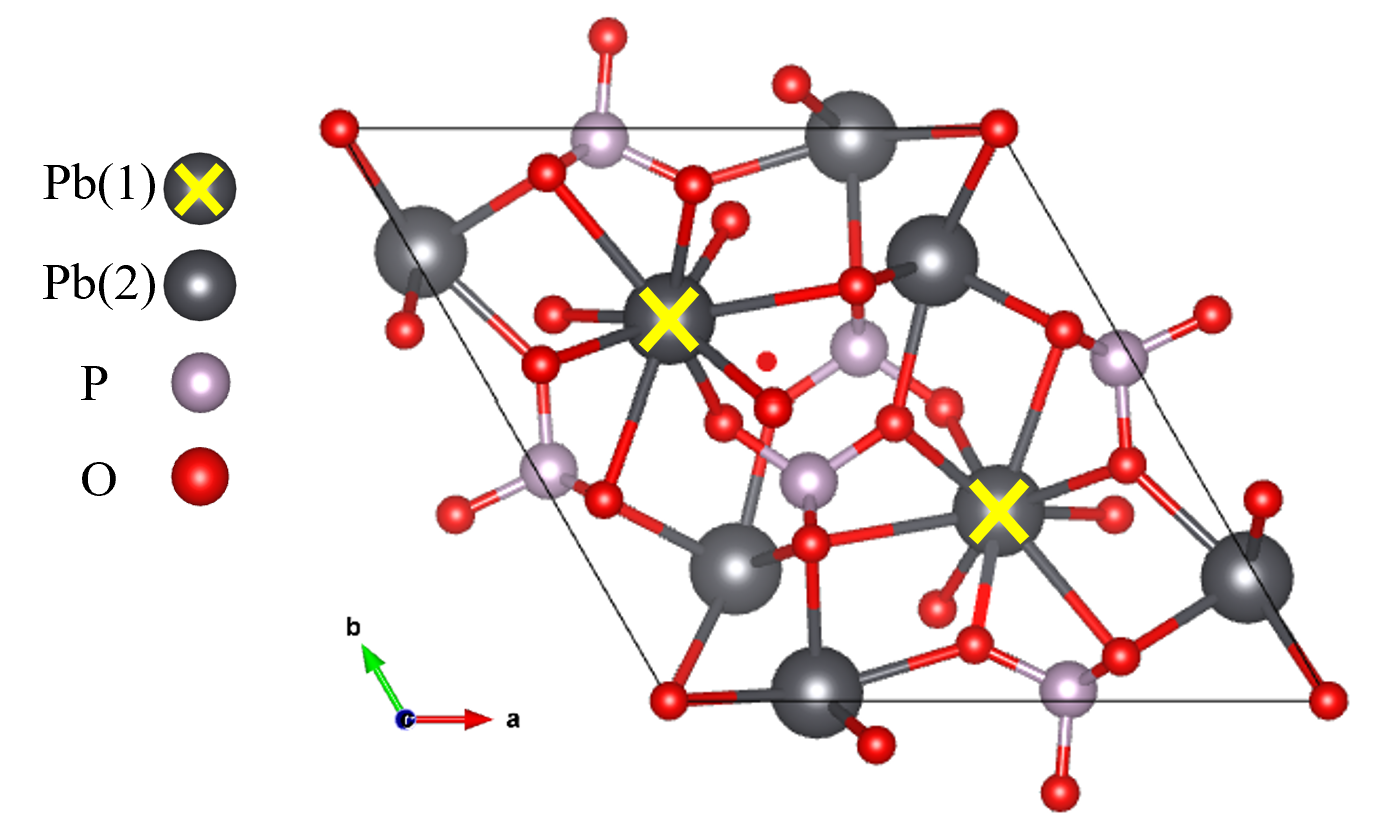}
    \caption{Crystal structure of Pb$_{10}$(PO$_{4}$)$_{6}$O$_{2}$ detailing two in-equivalent Pb sites.}
    \label{fig:LK99}
\end{figure}

\begin{figure*}
    \centering
    \includegraphics[width=8cm]{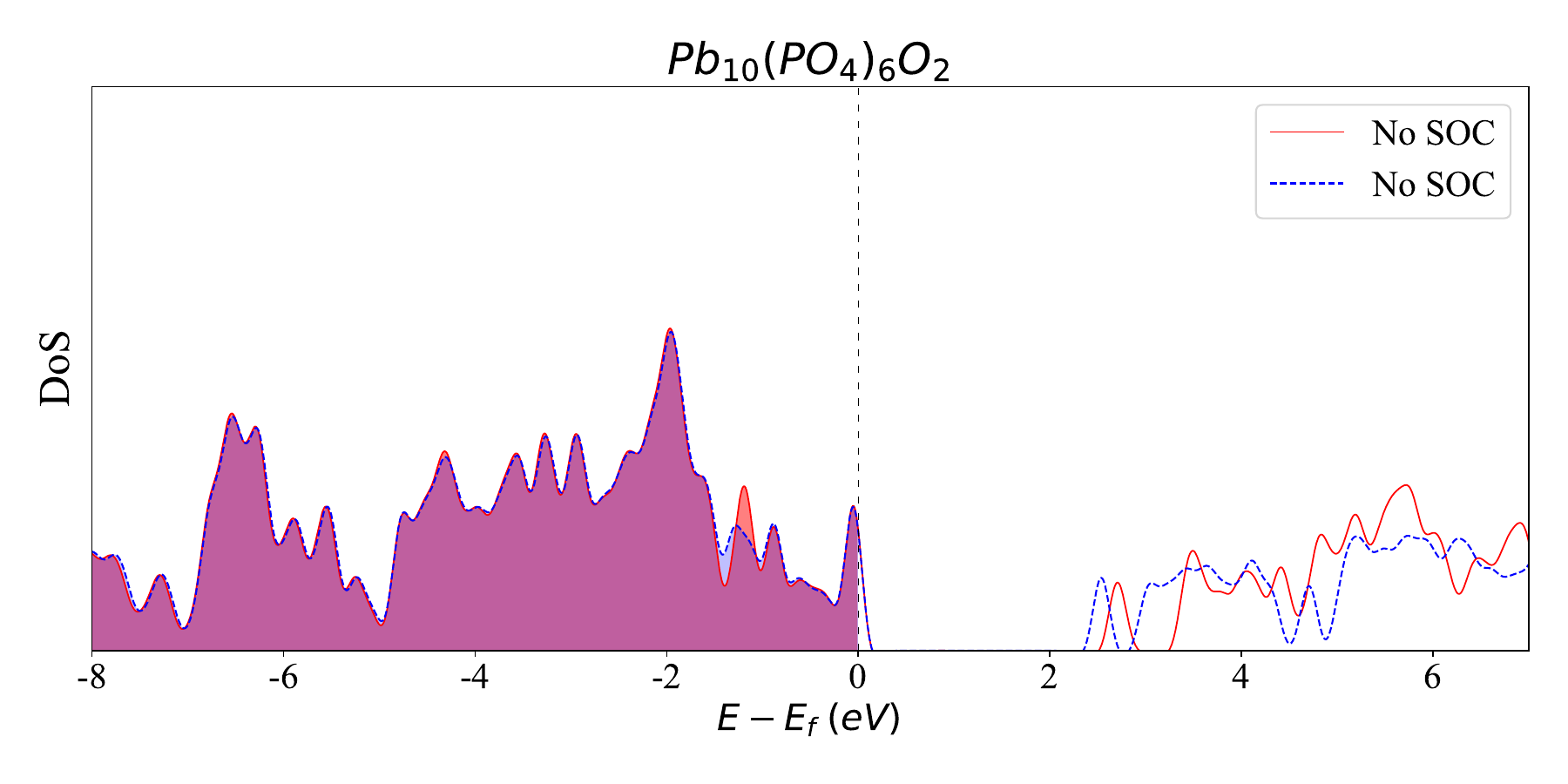}
    \includegraphics[width=8cm]{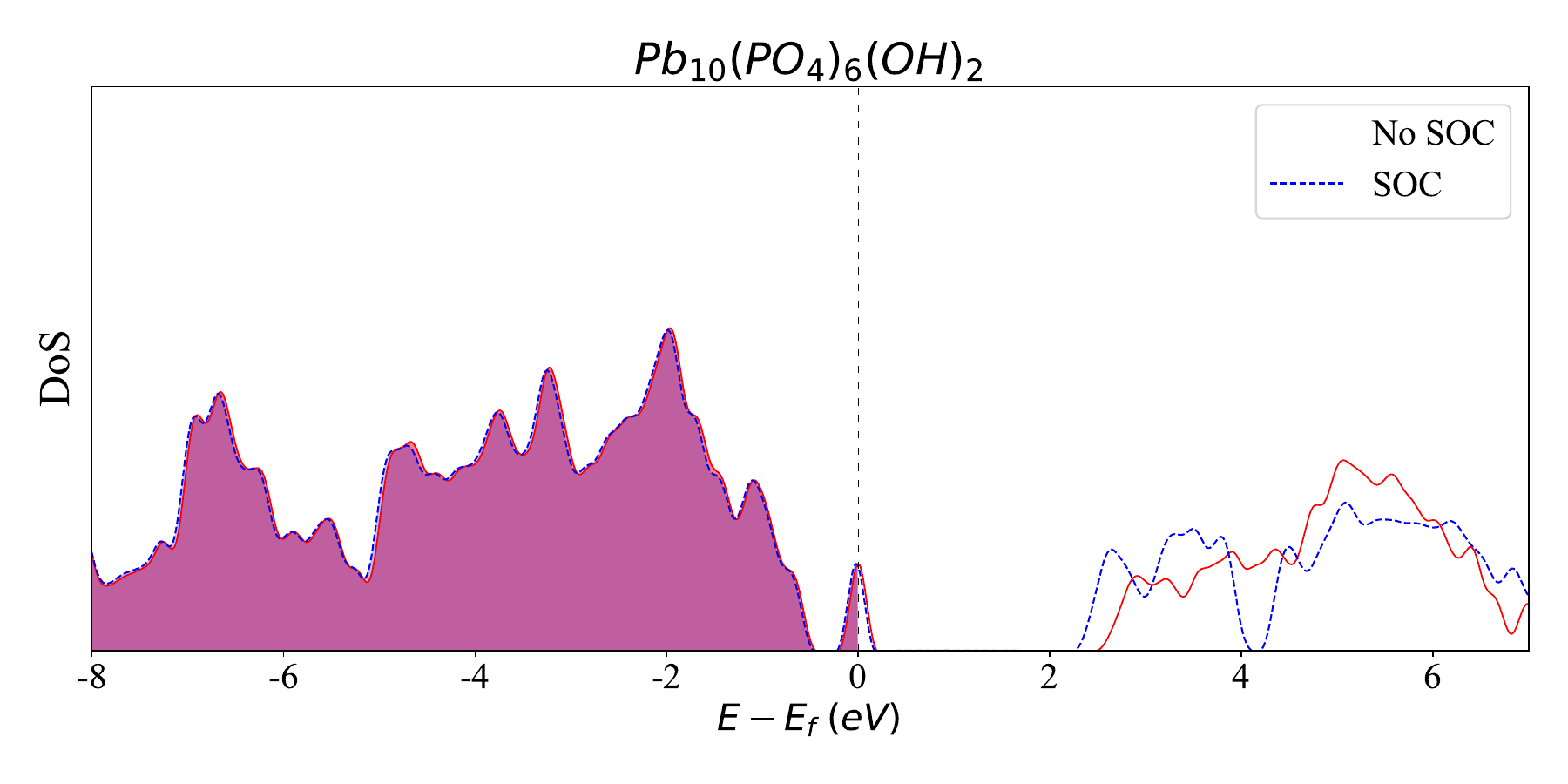}
    \caption{{\color{black}Density of states for Pb$_{9}$Cu(PO$_{4}$)$_{6}$X$_{2}$ for (a) X=0 and (b) X=OH. In each case the density of states is computed in the presence and absence of spin-orbit coupling.}}
    \label{fig:LK99dos}
\end{figure*}

\par 
{\color{black}In this work we follow a recently established protocol to compute a proxy for the electron-phonon coupling strength, $\lambda$. This proxy involves computation of the electron-phonon coupling at the $\Gamma$ point only, allowing us to overcome the computational challenges that have kept these quantities absent from the literature.} In Ref. \cite{PhysRevMaterials.6.074801}, it is shown that,
\begin{equation}
    \lambda\approx f \lambda_{\Gamma}
\end{equation} where it is reasonable to estimate $f$ as a constant for a family of materials. It was further shown in Ref. \cite{PhysRevMaterials.6.074801}, that the magnitude of $\lambda_{\Gamma}$ is generally sufficient for distinguishing systems likely to support superconductivity from those in which it is highly unlikely.  As a result, we have computed $\lambda_{\Gamma}$ for Pb$_{10}$(PO$_{4}$)$_{6}$X$_{2}$ X=O, OH, with copper substitution at the Pb(1) site, allowing for a direct comparison between the compositions. 
\begin{figure*}[t]
\subfigure[]{
\includegraphics[scale=0.3]{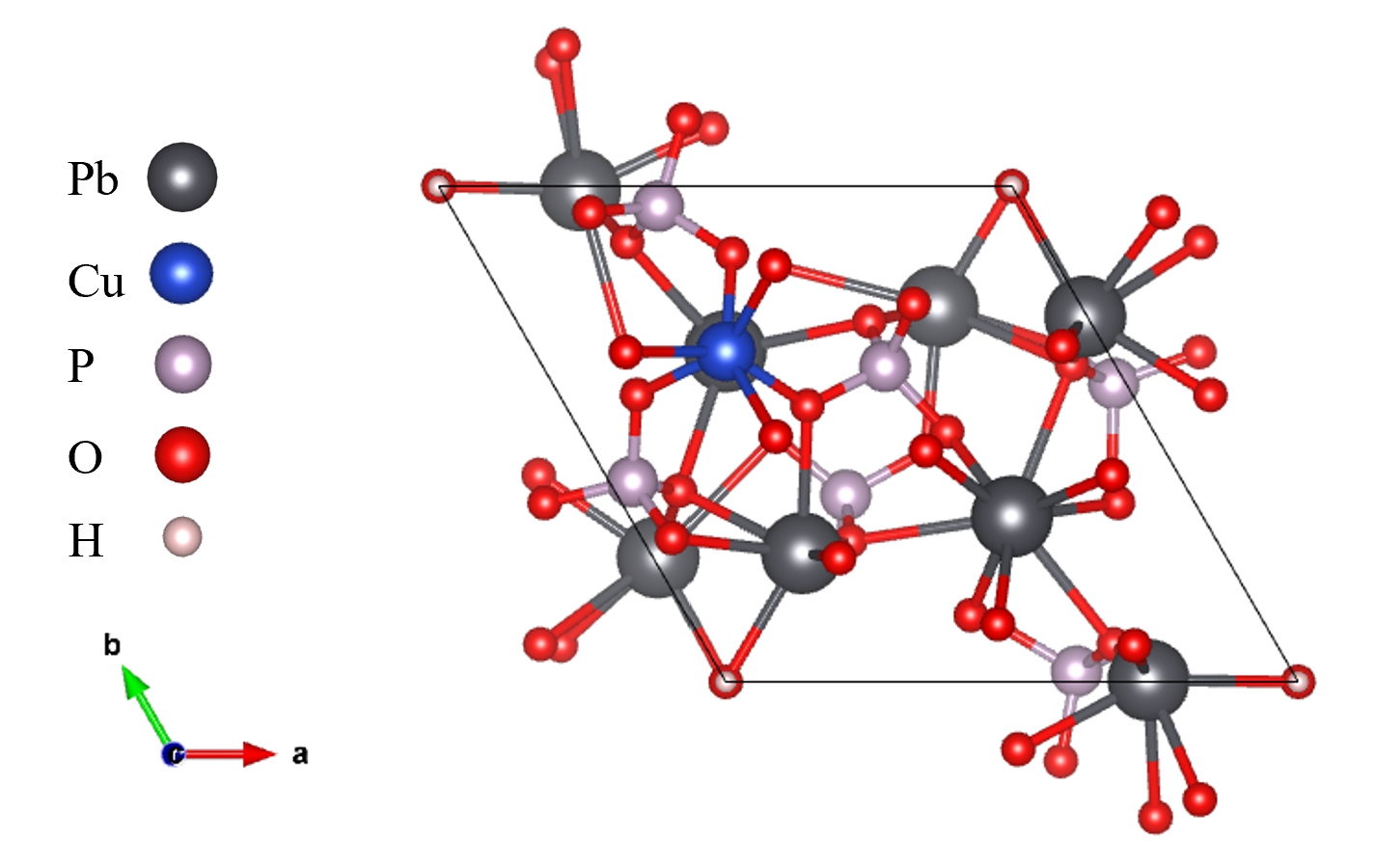}
\label{fig:}}
\subfigure[]{
\includegraphics[scale=0.21]{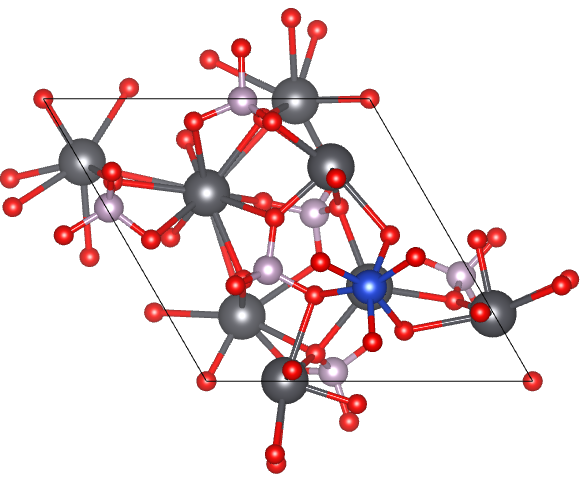}
\label{fig:}}
\subfigure[]{
\includegraphics[scale=0.28]{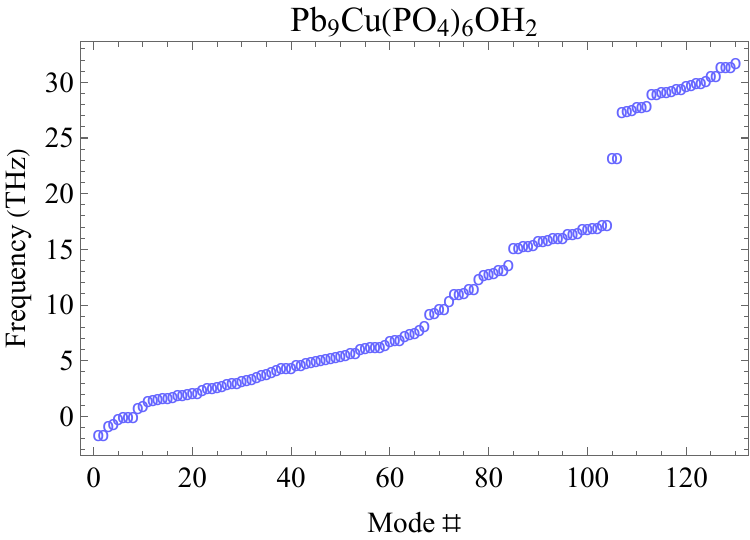}
\label{fig:}}
\subfigure[]{
\includegraphics[scale=0.28]{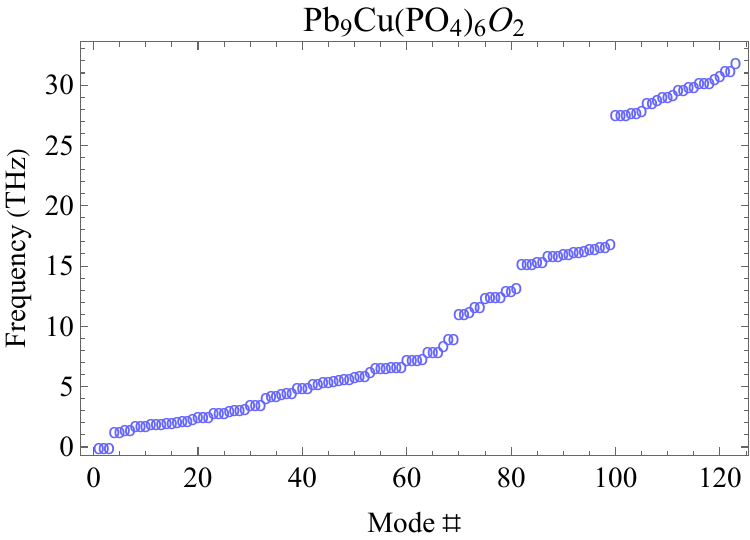}
\label{fig:}}
\caption{Relaxed crystal structure for Pb$_{9}$Cu(PO$_{4}$)$_{6}$X$_{2}$ with Cu substitution at Pb(1) for (a) X=OH and (b) X=O. The computed phonon modes at the $\Gamma$ location for X=OH, O are shown in (c) and (d) respectively. }
\label{fig:Compounds}
\end{figure*}
\par 
Our work finds that the magnitude of $\lambda_{\Gamma}$ is small and generally equivalent among the distinct compositions, making the prospect of high-temperature superconductivity by means of the electron-phonon interaction exceedingly unlikely for each composition. Nevertheless, low-temperature superconductivity remains possible.  

\section{Results}

In this work, all first principles calculations based on density-functional theory (DFT) are carried out using the Quantum Espresso software package \cite{QE-2009,QE-2017,QE-2020}. Exchange-correlation potentials use the Perdew-Burke-Ernzerhof (PBE) parametrization of the generalized gradient approximation (GGA) \cite{Perdew1996}. In each case the atomic positions and lattice parameters are taken from Ref. \cite{griffin2023origin}. The atomic positions and lattice parameters are subsequently relaxed until the maximum force on each atom is less than $1\times 10^{-5} \text{Ry/Bohr}$. The electron-phonon coupling strength is determined using density function perturbation theory (DFPT) as implemented in the Quantum Espresso software package. Given that LK99 is proposed to support superconductivity at room temperature, spin-orbit coupling is neglected. A 8 x 8 x 6 Monkhorst-Pack grid of k-points is utilized as well as a plane wave cutoff of 520 eV.  We have neglected inclusion of a Hubbard U for copper in these computations, which can admittedly impact the computed values. 

\begin{figure}
    \centering
    \includegraphics[width=8cm]{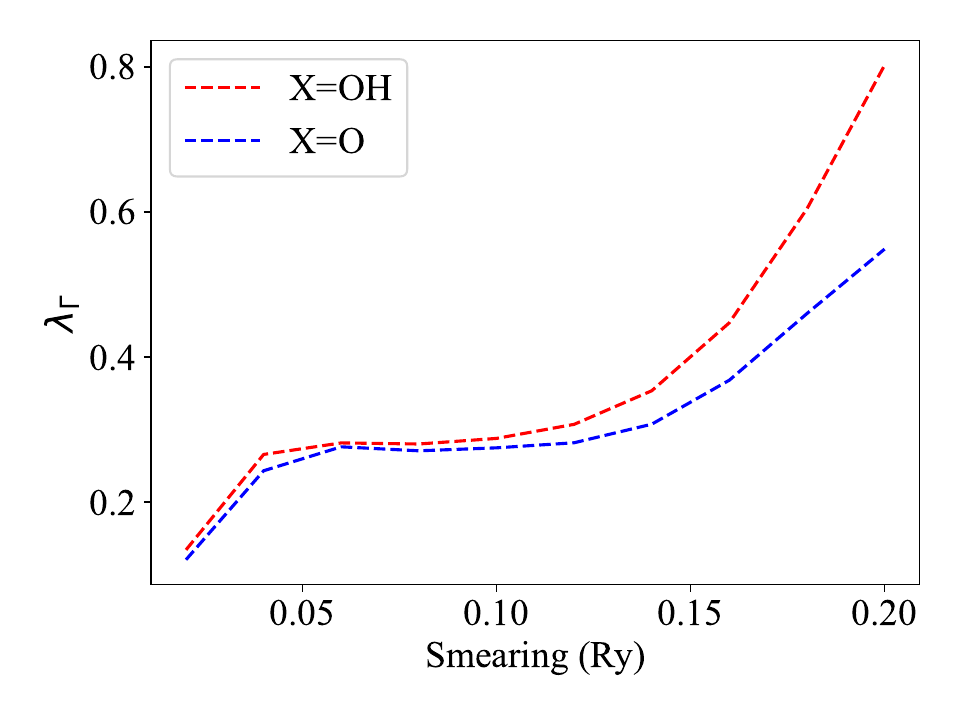}
    \caption{Electron-phonon coupling at the $\Gamma$ location and Fermi energy for copper substituted at the Pb(1) site in Pb$_{9}$Cu(PO$_{4}$)$_{6}$X$_{2}$ with X=O,OH, as a function of the smearing value used.}
    \label{fig:ElPh}
\end{figure}
\par 
Pb$_{10}$(PO$_{4}$)$_{6}$X$_{2}$ belongs to space group $P6_{3}/m$ and $P6_{3}$ for X=O, OH respectively. Details of the structure before and after copper substitution can be found in Ref. \cite{griffin2023origin}. The locations for copper substitution on the two in-equivalent Pb sites, labeled Pb(1) and Pb(2) are shown in Fig. \ref{fig:LK99}. {\color{black}The computed density of states for the relaxed structures for each of the compounds after copper substitution can be seen in Fig. \ref{fig:LK99dos} and the relaxed crystal structures are visible in Fig. \ref{fig:Compounds}. The density of states computations in Fig. \ref{fig:LK99dos} demonstrate that the inclusion of spin-orbit coupling has had a minimal impact below the Fermi energy, indicating that neglecting spin-orbit coupling will not invalidate the results obtained.} For further information regarding the crystal structure, bulk band structure, and more, we refer the reader to Refs. \cite{griffin2023origin,cabezas2023theoretical,shen2023phase,jiang2308pb9cu}.
\par 
The resulting phonon modes at the $\Gamma$ location are shown for Pb$_{9}$Cu(PO$_{4}$)$_{6}$X$_{2}$ with Cu at the Pb(1) site for X=OH, O in Fig. \ref{fig:Compounds}. We immediately note that the absence of negative frequency modes for X=O, indicating that this compound is dynamically stable. In contrast, for X=OH negative phonon modes can be found. While this can indicate a dynamic instability, it is likely that such negative frequency modes would be removed through imposition of a stricter convergence threshold when relaxing the structure. This presents a computational challenge due to the lightness of the hydrogen atoms whose positions can be easily adjusted with minimal effects on the energy of the structure, slowing convergence.   
\par 
The computed electron-phonon coupling strength at the $\Gamma$ location at the Fermi energy as a function of the smearing values utilized in the computation is shown in Fig. \ref{fig:ElPh}. In order to estimate $\lambda$ for each compound, we utilize the value obtained in Ref. \cite{paudyal2023implications} for Pb$_{9}$Cu(PO$_{4}$)$_{6}$O$_{2}$ with Cu substituted at the Pb(2) site. We then determine $f$, where 
\begin{equation}
    f=\lambda_{\Gamma}/\lambda\approx 3.5.
\end{equation} 

\par
The value of $\lambda_{\Gamma}$ used in this computation is determined by computing the electron-phonon coupling at $\Gamma$ for the same compound in Ref. \cite{paudyal2023implications} at each smearing value seen in Fig. \ref{fig:ElPh}. Using these results, we determine $\lambda$ for Cu substituted at the Pb(1) site with X=OH and X=O respectively at the finest smearing value, placing the results in Tab. \ref{tab:1}.
\begin{table}[t]

\begin{center}
\begin{tabular}{ |m{2cm}|m{2cm}|m{2cm}| } 
 \hline
 X & $\lambda_{\Gamma}$ & $\lambda$ \\ 
 \hline
 O & 0.120 &  0.43\\ 
 \hline
 OH & 0.134 & 0.46 \\ 
 \hline
\end{tabular}
\caption{Estimates of $\lambda$ for Cu substituted at the Pb(1) site with X=OH and X=O respectively utilizing the finest value smearing value for computing density of states at the Fermi energy.}
\label{tab:1}
\end{center}
\end{table}

\par 
While not small enough to rule out low-temperature superconductivity, the computed values of electron-phonon coupling are weak in comparison to known conventional, high-temperature superconductors such as the high-pressure hydrates, where it is possible to obtain $\lambda\approx 6$. Hence we assume that electron-phonon coupling in all the compositions we considered is not large enough to account for possible high Tc superconducting state. 

In Ref. \cite{paudyal2023implications}, a superconducting critical temperature of $<2K$ was computed for Pb$_{9}$Cu(PO$_{4}$)$_{6}$O$_{2}$ with Cu substituted at the Pb(2) site. Given the current results, a similar value should be obtained for substitution at the Pb(1) site.

\section{Conclusion}
\par 
The approximated electron-phonon coupling is comparably small for copper substitution at the Pb(1) and Pb(2) sites, indicating that if LK99 is a candidate for high-temperature superconductivity, it is highly unlikely that electron-phonon coupling is the mechanism, but this is a possible mechanism for low-temperature superconductivity.

This conclusion is in agreement with the current literature which has been unable to achieve the proposed claim of room-temperature, ambient pressure superconductivity. Furthermore, there exists evidence that magnetic configurations not considered in this study may be stable and further limit the possibility of high-temperature superconductivity.  

We hope that experimental data of electron-phonon coupling in LK99 materials will be available soon and provide context for the values predicted in this work.
\par

\acknowledgements{}

We are grateful to G. Aeppli for discussions. We acknowledge support from the European Research Council under the European Union Seventh Framework ERS-2018-SYG 810451 HERO and the University of Connecticut.

\bibliography{ref.bib}
\end{document}